\documentclass[letterpaper, 10 pt, conference]{ieeeconf}  

\IEEEoverridecommandlockouts                              

\usepackage[utf8]{inputenc} 
\usepackage[T1]{fontenc}    
\usepackage{hyperref}       
\usepackage{url}            
\usepackage{booktabs}       
\usepackage{amsfonts}       
\usepackage{nicefrac}       
\usepackage{microtype}      
\usepackage{xcolor}         
\usepackage{amsmath}
\usepackage{subcaption}
\usepackage{graphicx, tikz}
\usepackage[ruled,vlined]{algorithm2e}
\usepackage{multirow}
\usepackage[font=scriptsize]{caption}
\usepackage{comment}
\usepackage{ulem}

\title{\LARGE \bf
Automatic Regression for Governing Equations with Control (ARGOSc)
}

\author{Amir~Bahador~Javadi,~\IEEEmembership{Graduate~Student~Member,~IEEE,}
        Amin~Kargarian,~\IEEEmembership{Senior~Member, IEEE, and}\\
        Mort Naraghi-Pour,~\IEEEmembership{Life~Senior~Member, IEEE}
\thanks{This material is based upon work supported by the National Science Foundation under Grant Number 1944752.}
\thanks{The authors are with the Department of Electrical and Computer Engineering,
Louisiana State University, Baton Rouge, LA, USA.
        {\tt\small ajavad2,kargarian,naraghi@lsu.edu}}
}

\begin{document}

\maketitle
\thispagestyle{empty}
\pagestyle{empty}

\begin{abstract}

Learning the governing equations of dynamical systems from data has drawn significant attention across diverse fields, including physics, engineering, robotics and control, economics, climate science, and healthcare. Sparse regression techniques, exemplified by the Automatic Regression for Governing Equations (ARGOS) framework, have demonstrated effectiveness in extracting parsimonious models from time series data. However, real-world dynamical systems are driven by input control, external forces, or human interventions, which standard ARGOS does not accommodate. To address this, we introduce ARGOS with control (ARGOSc), an extension of ARGOS that incorporates external control inputs into the system identification process. ARGOSc extends the sparse regression framework to infer governing equations while accounting for the effects of exogenous inputs, enabling robust identification of forcing dynamics in low- to medium-noise datasets. We demonstrate ARGOSc’s efficacy on benchmark systems, including the Van der Pol oscillator, Lotka-Volterra, and the Lorenz system with forcing and feedback control, showing enhanced accuracy in discovering governing laws. \uline{Under the noisy conditions, ARGOSc outperforms the widely used sparse identification of nonlinear dynamics with control (SINDYc), in accurately identifying the underlying forced dynamics. In some cases, SINDYc fails to capture the true system dynamics, whereas ARGOSc consistently succeeds.}

\end{abstract}


\section{INTRODUCTION}

Learning governing equations from data is essential, yet remains a fundamental challenge in dynamical system modeling, with wide-ranging applications across fields such as physics and engineering \cite{physics}, robotics and control \cite{control}, economics \cite{eco}, climate science \cite{climate}, and healthcare \cite{health}. Traditional methods for deriving mathematical representations of system dynamics often rely on expert knowledge and manual parameter tuning, making them inefficient for large, complex systems \cite{chen}. In recent years, data-driven artificial intelligence (AI)-based approaches such as sparse regression have enabled the automatic discovery of differential equations from empirical observations, reducing reliance on prior assumptions \cite{brunton2016discovering}. Sparse identification of nonlinear dynamical systems (SINDy) is a data-driven method widely adopted across various fields to discover simplified mathematical equations representing complex dynamical systems. It identifies a sparse set of governing equations from observed data by promoting simplicity and interpretability. The term "sparse" refers to selecting only a few essential terms, making the resulting models efficient and easy to analyze. SINDy's popularity stems from its ability to effectively balance model complexity and accuracy, thus making it an attractive choice for researchers aiming to uncover underlying dynamics from experimental or observational data. However, SINDy faces limitations when dealing with noisy data and requires careful hyperparameter tuning \cite{hirsh2022sparsifying}. In addition, SINDy heavily depends on accurately selecting basis functions to capture nonlinear dynamics \cite{brunton2016discovering}. To address these challenges, SINDy has been expanded and improved with methods such as Ensemble-SINDy \cite{ensemble} and SINDy with the Akaike Information Criterion (AIC) \cite{sindy-aic}. However, as demonstrated in \cite{Egan2024}, these methods have a lower success rate and reduced robustness than automatic regression for governing equations (ARGOS), particularly under noisy conditions.

In 2024, ARGOS was introduced as a framework that integrates statistical learning techniques to extract differential equations from observational data \cite{Egan2024}. ARGOS enhances traditional sparse regression methods by incorporating signal denoising, bootstrapped confidence intervals for uncertainty quantification, and adaptive model selection. By automating the identification process, ARGOS improves accuracy in discovering dynamical laws, particularly in high-dimensional, noisy environments where conventional approaches struggle.

A crucial aspect often overlooked in conventional system identification learning frameworks is the presence of \textbf {\textit{control inputs}}, \textbf{\textit{external forces}}, or \textbf{\textit{human interventions}}, which play a fundamental role in real-world dynamical systems \cite{kaiser2018sparse}. Many physical and engineering systems, including power grids \cite{power}, robotic systems \cite{robotic}, and biological networks \cite{bio}, operate under external actions that directly influence their dynamics \cite{yuan}. Ignoring these inputs can lead to incomplete or inaccurate models, limiting their applicability in practical settings. Therefore, it is imperative to develop methodologies that explicitly incorporate these external forces into the equation learning process, ensuring that the identified models accurately reflect the actual governing dynamics. To address this challenge, SINDYc, an extension of SINDy that considers external forcing, was introduced in \cite{sindyc} to improve the identification of underlying dynamics. As SINDYc has been built within the same framework as SINDy, it carries the same limitations. Therefore, to address these challenges and build on the advantages of ARGOS, extending ARGOS to incorporate control mechanisms can enhance its robustness and generalization capabilities beyond ARGOS alone.

In this work, we present ARGOS with control (ARGOSc), a novel learning approach extending ARGOS to directly incorporate control signals into the design matrix. ARGOSc enables a more accurate and interpretable representation of dynamical systems under external influence by augmenting the regression process with forcing-dependent terms. By explicitly incorporating forcing inputs into the discovery framework, ARGOS extends its applicability to dynamical systems with control inputs, external forces, or human interventions, addressing a critical gap in existing techniques. We validate the approach on well-known dynamical systems, demonstrating its ability to consistently recover governing equations with minimal human intervention.

By advancing automation of model discovery and integrating forcing inputs into the identification process, ARGOSc presents a significant step forward in system identification. This enhancement is particularly relevant for applications in power systems, robotics, and other domains where forced dynamical systems are prevalent. Integrating statistical inference with sparse regression enables a more reliable and interpretable extraction of differential equations, paving the way for enhanced predictive modeling and control strategies.

\textbf{Remark.} In the context of learning equations of dynamical systems, the terms \textit{control inputs}, \textit{external forces}, and \textit{human intervention} can sometimes be used interchangeably depending on how they influence system behavior. For example, in control systems, an external force applied to a system can be modeled as an input control when it is systematically designed to influence system behavior \cite{ogata}. Likewise, human intervention can function as an external force when an operator manually adjusts control variables, effectively acting as a dynamic input \cite{astrom}. Thus, in this paper, we use these terms interchangeably.

\textbf{Contributions.}
\label{contribute}
In this work, we extend the ARGOS framework to incorporate external control inputs into system identification. Our main contributions include:
\begin{itemize}
\item We propose an ARGOS framework extension that directly integrates control inputs into the design matrix, enabling more accurate discovery of governing equations in forced dynamical systems.
\item We enhance the identification of dynamical systems subjected to external forcing by modifying the design matrix to incorporate both state-dependent and control-dependent terms, thereby achieving improved generalization compared to ARGOS.
 
\item We validate ARGOSc on well-known nonlinear systems, including the Van der Pol oscillator, Lotka-Volterra system, and the Lorenz system with control, demonstrating its effectiveness in accurately capturing dynamics under medium to low noise levels.
\item We provide a comparative analysis with SINDYc and standard ARGOS, showing that ARGOSc outperforms existing methods in accurately identifying governing equations, particularly in scenarios involving external control signals.
\item \uline{We demonstrated that ARGOSc outperforms ARGOS and SINDYc in capturing underlying dynamics, and highlighted SINDYc’s failures in Lotka–Volterra and Lorenz case studies, where ARGOSc succeeds.}
\end{itemize}

\section{Preliminaries and proposed method}
\label{method}
\subsection{ARGOS}
\begin{figure}[h]
\centering
\resizebox{0.5\textwidth}{!}{
\begin{tikzpicture}[
    block/.style = {draw, rectangle, minimum height=1cm, minimum width=3cm, align=center},
    line/.style = {draw, -},
    arrow/.style = {line width=0.75mm, draw, -{latex}},
    node distance = 0.8cm and 0.8cm
]
\usetikzlibrary{positioning}

\node[block] (input) {\large Real/synthetic data};
\node[block, right=of input] (filter) {\large Savitzky-Golay \\ \large filter};
\node[block, right=of filter, fill=gray!30] (sparse) {\large Sparse regression};
\node[block, right=of sparse] (boot) {\large Bootstrap sampling \& \\ \large confidence intervals};
\node[block, right=of boot] (output) {\large Learned dynamics};

\draw[arrow] (input) -- (filter);
\draw[arrow] (filter) -- (sparse);
\draw[arrow] (sparse) -- (boot);
\draw[arrow] (boot) -- (output);

\end{tikzpicture}
}
\caption{Flowchart of the ARGOS methodology for identifying governing equations.}
\label{schematic}
\end{figure}
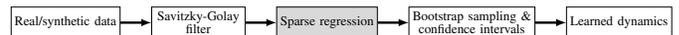

To implement ARGOS, as illustrated in Fig.\ref{schematic}, the first step after collecting real or synthetic data is to apply the Savitzky-Golay filter \cite{savitzky}. This filter fits a polynomial over a local window of data points, reducing noise while preserving the underlying signal characteristics. Then, numerical derivatives of the time-series data are computed using the smoothed signal \cite{Egan2024}. After smoothing and differentiating the data, a design matrix is constructed with monomials up to the \textit{d}-th degree. The regression problem is formulated using this design matrix, and the least absolute shrinkage and selection operator (LASSO) \cite{statis} or adaptive LASSO \cite{alasso} is applied during model selection to shrink coefficients and select non-zero terms. After an initial sparse regression estimate is obtained, the design matrix is trimmed to include only the highest-order variables with non-zero coefficients, and sparse regression is reapplied. A grid of thresholds is used to select models based on coefficients that exceed specific thresholds. Ordinary least squares (OLS) is performed on the selected variables to compute unbiased coefficients, and the model with the minimum Bayesian information criterion (BIC) is chosen. Bootstrapping is applied to generate 2000 sample estimates, and 95\% bootstrap confidence intervals are constructed to identify the final model, including only variables whose confidence intervals do not include zero and whose point estimates fall within the intervals \cite{Egan2024}. 

To extend the ARGOS framework to include control, denoted as ARGOS with control (ARGOSc), we propose to modify the sparse regression block (the gray box in Fig.\ref{schematic}), as detailed in section \ref{argosc}.

\subsection{ARGOSc}
\label{argosc} Consider a dynamical system with forcing inputs. Such systems are typically modeled within the framework of ordinary differential equations:
\begin{align}
    \begin{cases}
    \dot{x}_j(t) = f_j(x(t), u(t)), \quad j = 1, \dots, m \\
    x(0) = x_0
\end{cases}
\end{align}
where  \( x(t) = [x_1(t), x_2(t), \dots, x_m(t)]^\top \in \mathbb{R}^m \), $u(t) \in \mathbb{R}^r$, and \( x_0 \in \mathbb{R}^m \) are the state vector, the forcing input, and the initial condition of the dynamical system, respectively. The function \( f_j(x(t), u(t)) \) characterizes the system's evolution over time, with the objective being to mathematically identify and explicitly represent this function. 

After smoothing and differentiation, a design matrix \(\Theta(X, U) \in \mathbb{R}^{n \times p}\) is constructed, containing candidate basis functions of state variables \( X \in \mathbb{R}^{n \times m} \) and forcing inputs \( U \in \mathbb{R}^{n \times r} \) up to a specified polynomial degree \(d\), where \( n \) is the number of time samples. Based on the nature and behavior of the underlying dynamics to be discovered by ARGOSc, additional basis functions can be added to the design matrix.

\begin{equation}
    \Theta(X,U) = 
    \begin{bmatrix}
    1 & X & \cdots & X^{[d]} & U & \cdots & U^{[d]} & \Phi(X,U)
    \end{bmatrix}
\end{equation}
where \( X^{[i]} \), and \( U^{[i]} \) are monomials of order \( i \) in \( x \), and monomials of order \( i \) in \( u \), respectively, and \(\Phi(X,U)\) includes nonlinear functions for enhanced representation (e.g., polynomials, trigonometrics, etc.). Finally, the linear regression problem is formulated as follows:
\begin{equation}
    \dot{X} = \Theta(X,U) B + E
    \label{eq}
\end{equation}
where \( \dot{X} \in \mathbb{R}^{n \times m} \) is the matrix of time derivatives, $B \in \mathbb{R}^{p \times m}$ (with \( p \) being the number of basis functions) and  $E \in \mathbb{R}^{n \times m}$ are coefficient and residual matrices (model errors), respectively. A sparse linear regression problem must be applied to each of the \(m\) states to discover the underlying dynamics of the forced dynamical system. Equation~\ref{eq} can be restated for each of the \(m\) states in the matrix \(\dot{X}\) as follows:

\begin{equation}
    \dot{x_j} = \Theta(X,U) \beta_j + \epsilon_j, \quad j=1,\dots,m
\end{equation}

where \( \beta_j \in \mathbb{R}^p \) is the coefficient vector corresponding to state \( x_j \), and \( \epsilon_j \in \mathbb{R}^n \) is the residual error vector for \( x_j \). To achieve this goal and discover a parsimonious model, either the LASSO \cite{statis} or adaptive LASSO \cite{alasso} can be applied. We aim to determine the values of the coefficient matrix B from the following,

\begin{equation}
    \hat{\beta}_j = \arg\min_{\beta_j} \left\{ \|\dot{x}_j - \Theta(X, U)\beta_j\|_2^2 + \lambda \sum_{i=1}^p \omega_{j,i}|\beta_{j,i}| \right\}
\end{equation}

where the regularization parameter $\lambda > 0$ is chosen to identify the Pareto-optimal model that achieves an optimal trade-off between model complexity and accuracy. The $\omega_{j,i}$ values represent the weighted penalties in adaptive LASSO. We have chosen adaptive LASSO since it assigns different penalty weights to different coefficients and thus can capture nonlinear dynamics more efficiently than LASSO.

We used the same algorithm described in the supplementary material of \cite{Egan2024} for smoothing and differentiating the data. However, we modify Algorithm 2 from the supplementary material of \cite{Egan2024} to account for external forcing inputs in addition to state variables, as presented in Algorithm \ref{alg1}.

\newcommand{\bbeta}{\boldsymbol{\beta}}
\newcommand{\bTheta}{\boldsymbol{\Theta}}
\newcommand{\bX}{\mathbf{X}}
\newcommand{\bU}{\mathbf{U}}
\newcommand{\dxj}{\dot{\mathbf{x}}_{j}}

\begin{algorithm}
\caption{ARGOSc}
\SetAlgoLined
\textbf{Input:} $\bX \in \mathbb{R}^{n \times m}$, $\bU \in \mathbb{R}^{n \times r}$, $\dxj \in \mathbb{R}^{n \times m}$, $d$, $\alpha = 0.05$. \\
\textbf{STEP ONE:} Initiate design matrix $p^{(0)} = \binom{m + d}{d}$\;  
Create $\bTheta^{(0)}(\bX, \bU) \in \mathbb{R}^{n \times p^{(0)}}$ using basis functions up to order $d$ of the columns of $\bX$\,, $\bU$ as well as any other customized functions if needed, based on the knowledge and/or the nature, and behavior of the dynamical system; \\ 
\textbf{STEP TWO:} Trim design matrix \\  
Variable selection with the lasso or adaptive lasso \\  
$\lambda^{*}$: Optimal $\lambda$ from 10-fold cross-validation \\  
lasso: $w = 1$ \\  
adaptive lasso: $w = \text{ridge regression coefficients}$ \\  
\[
\hat{\beta}^{(0)} = \underset{\beta}{\arg\min} \left\| \dxj - \bTheta^{(0)}(\bX, \bU) \beta \right\|_2^2 + \lambda^{*} \sum_{k=1}^{p^{(0)}} w_k |\beta_k|;
\]  
Extract $\bTheta^{(1)}(\bX, \bU)$ to contain columns of $\bTheta^{(0)}(\bX, \bU)$ up to the largest order $d^{(1)}$ of the selected variables in $\hat{\beta}^{(0)}$\ as
well as any other customized functions; \\ 
\textbf{STEP THREE:} Final point estimates \\  
Repeat sparse regression algorithm from STEP TWO \\  
$p^{(1)} = \binom{m + d^{(1)}}{d^{(1)}}$\;  
\[
\hat{\beta}^{(1)} = \underset{\beta}{\arg\min} \left\| \dxj - \bTheta^{(1)}(\bX, \bU) \beta \right\|_2^2 + \lambda^{*} \sum_{k=1}^{p^{(1)}} w_k |\beta_k|
\]  
Apply threshold values \\  
$\eta = [10^{-8}, 10^{-7}, \ldots, 10^{1}]$\;  
\For{$i = 1, \ldots, \text{card}(\eta)$}{  
  Ordinary least squares regression (OLS) estimate after variable selection:  
\[
\begin{aligned}
\hat{\beta}^{\text{OLS}}[i] &= \underset{\beta_{\mathcal{K}_i}}{\arg\min} 
\left\| \dxj - \bTheta_{\mathcal{K}_i}^{(1)}(\bX, \bU) \beta_{\mathcal{K}_i} \right\|_2^2 \\
&\text{where } \mathcal{K}_i = \{k : |\hat{\beta}_k^{(1)}| \geq \eta_i \}
\end{aligned}
\]
\[
  \text{BIC}_i = \text{BIC}(\hat{\beta}^{\text{OLS}}[i])
\]  
}  
$\hat{\beta} = \left\{ \hat{\beta}^{\text{OLS}}[i] \Big| i : \underset{i}{\arg\min}\,(\text{BIC}) \right\} $\;  
\textbf{STEP FOUR:} Bootstrap estimates for confidence intervals \\  
$B = 2000$ bootstrap samples\\
bootstrap Statements 6--10 to approximate confidence interval bounds: $CI_{\text{lo}} = [B\alpha/2]$ and $CI_{\text{up}} = B - CI_{\text{lo}} + 1$\;  
Construct bootstrap confidence intervals for $\hat{\beta}$:  
\[
\begin{aligned}
\hat{\beta}_k &\in \left[ \hat{\beta}_k^{\text{OLS}\{CI_{\text{lo}}\}}, \hat{\beta}_k^{\text{OLS}\{CI_{\text{up}}\}} \right], \\
&\text{and } \quad 0 < \hat{\beta}_k^{\text{OLS}\{CI_{\text{lo}}\}} 
\ \text{or} \ 
0 > \hat{\beta}_k^{\text{OLS}\{CI_{\text{up}}\}}
\end{aligned}
\]
\label{alg1}
\end{algorithm}

We include these assumptions to ensure the model remains tractable and representative of real-world systems. 

\textbf{Assumption 1} (Model sparsity). The sparsity assumption posits that the governing equations of a dynamical system can be represented with only a few active terms from a potentially large set of candidate functions. This aligns with the principle of Occam’s Razor, where the simplest possible model that accurately describes the data is preferred. Sparse modeling is particularly useful in real-world applications where physical systems often follow concise governing equations. 



\textbf{Assumption 2} (Presence of governing terms in design matrix). This assumption states that the actual governing terms of the dynamical system must be included in the design matrix. The effectiveness of sparse regression depends on the completeness of the candidate design matrix. If the actual governing terms are absent, sparse regression methods may yield an inaccurate or incomplete model. Models lacking key terms fail to capture system dynamics, especially when noise levels are high. 



\section{Dynamical System Benchmarks}
\label{benchmark}
Most dynamical systems can be formulated as a set of ordinary differential equations that describe the evolution of state variables over time. Classic examples include the Van der Pol oscillator, which models nonlinear electrical circuits and biological rhythms, the Lotka-Volterra equations, which describe predator-prey interactions in ecological systems, and the Lorenz system, a foundational model in chaos theory and atmospheric dynamics.
\subsection{Van der Pol Oscillator}
The Van der Pol oscillator is a nonlinear second-order system that models self-sustained oscillations in circuits, biological rhythms, and neural activity. The feedback control system is defined as:
\begin{align}
\begin{cases}
\dot{x}_1 = x_2, \\
\dot{x}_2 = \mu(1 - x_1^2)x_2 - x_1 + u(t), \\
u(t) = -k_p x_1 - k_i x_2.
\end{cases}
\end{align}
where \( x_1 \) and \( x_2 \) denote position and velocity, respectively, with \( \mu = 1.2 \) governing nonlinearity. The feedback control law, defined by proportional (\( k_p = 1 \)) and integral (\( k_i = 1 \)) gains, stabilizes the system by regulating both states to achieve the desired behavior.

\subsection{Lotka-Volterra System}
The Lotka-Volterra system consists of first-order nonlinear differential equations modeling predator-prey dynamics. The forced system is given by:
\begin{align}
\begin{cases}
\dot{x}_1 = x_1(a - b x_2) + u(t), \\
\dot{x}_2 = -x_2(c - d x_1), \\
u(t) = k_u \sin(t).
\end{cases}
\end{align}
where \( x_1 \) and \( x_2 \) represent prey and predator populations, respectively, with parameters \( a = 8, b = 1, c = 4, d = 1 \) defining interaction rates. The control gain (\( k_u = 1 \)) influences prey dynamics to regulate or stabilize the system’s behavior.

\subsection{Lorenz System}
The Lorenz system is a nonlinear model for chaotic dynamics, often representing atmospheric convection and complex systems. The system with forcing input is given by:
\begin{align}
\begin{cases}
\dot{x}_1 = \sigma (x_2 - x_1) + u(t), \\
\dot{x}_2 = x_1(\rho - x_3) - x_2, \\
\dot{x}_3 = x_1 x_2 - \beta x_3, \\
u(t) = k_u \cos^3(t).
\end{cases}
\end{align}
where \( x_1, x_2, x_3 \) are state variables, and \( \sigma = 10, \rho = 28, \beta = \frac{8}{3} \) define system behavior. The control gain (\( k_u = 1 \)) introduces periodic forcing to influence the system’s chaotic dynamics, potentially stabilizing or altering its behavior.

\section{Numerical Experiments}
\label{expreiment}
Our primary focus is to compare the performance of ARGOSc with its widely used competitor, SINDYc. We conduct experiments on various dynamical system test cases, each simulated over 30 seconds under different noise conditions. For each case, the first 10 seconds of simulated data, sampled at a time step of 0.001 s, provide 10,000 training samples. Gaussian noise with varying signal-to-noise ratios is added to the generated dataset for each system to evaluate model robustness: 25 dB and 14 dB for the Van der Pol oscillator and the Lotka-Volterra system, and 49 dB and 37 dB for the Lorenz system. The trained models are then validated using the remaining 20 seconds of data to assess predictive accuracy. It should be noted that all simulations were conducted on a PC equipped with an Intel(R) Xeon(R) CPU E5-1630 v3 @ 3.70 GHz and 16.0 GB of RAM.

To highlight the limitations of ARGOS in isolation, this method is applied exclusively to the Lorenz system alongside the other two models. The same training process is also used to assess its ability to capture the forced dynamical system dataset after training on the testing dataset.

\subsection{Van der Pol Oscillator} 
\begin{figure}
\vspace{5pt}
    \centering
    \begin{subfigure}[b]{0.5\textwidth}
        \centering
        \begin{subfigure}{0.48\textwidth}
            \centering
            \includegraphics[width=\linewidth]{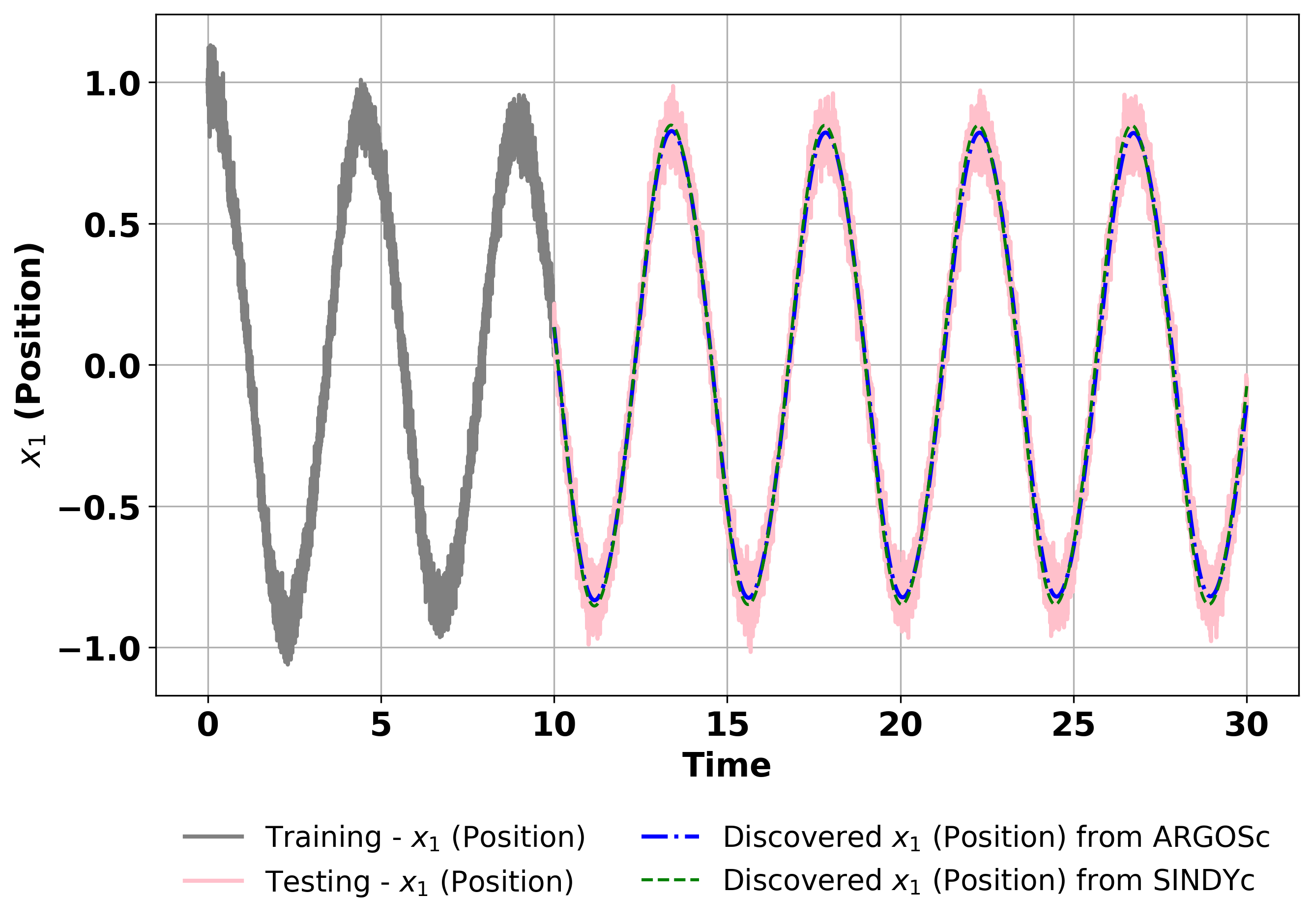}
            \label{x1_25db}
        \end{subfigure}
        \hfill
        \begin{subfigure}{0.48\textwidth}
            \centering
            \includegraphics[width=\linewidth]{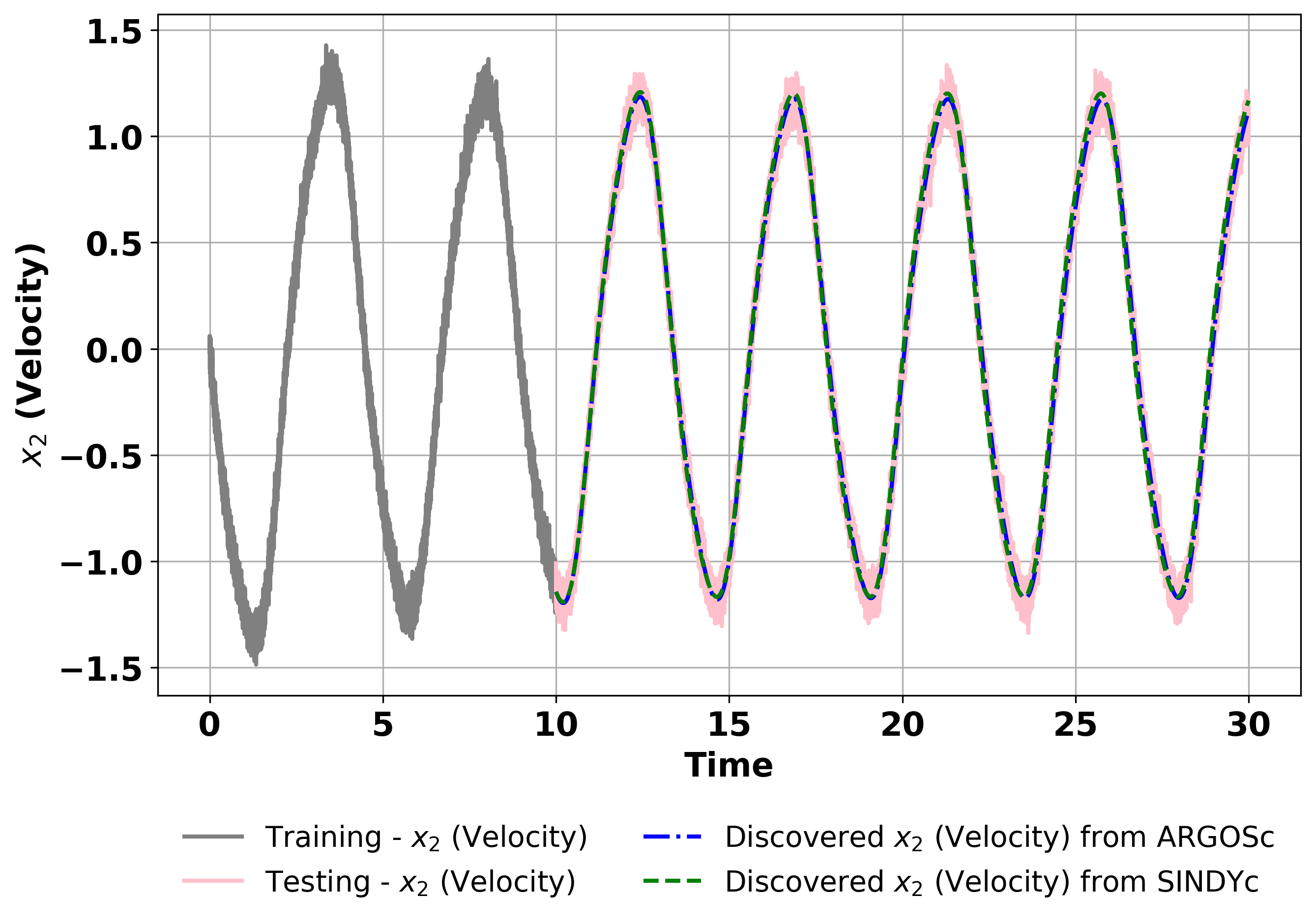}
            \label{x2_25db}
        \end{subfigure}
        \vspace{-10pt}
        \subcaption{SNR = 25 dB}
    \end{subfigure}

    \begin{subfigure}[b]{0.5\textwidth}
        \centering
        \begin{subfigure}{0.48\textwidth}
            \centering
            \includegraphics[width=\linewidth]{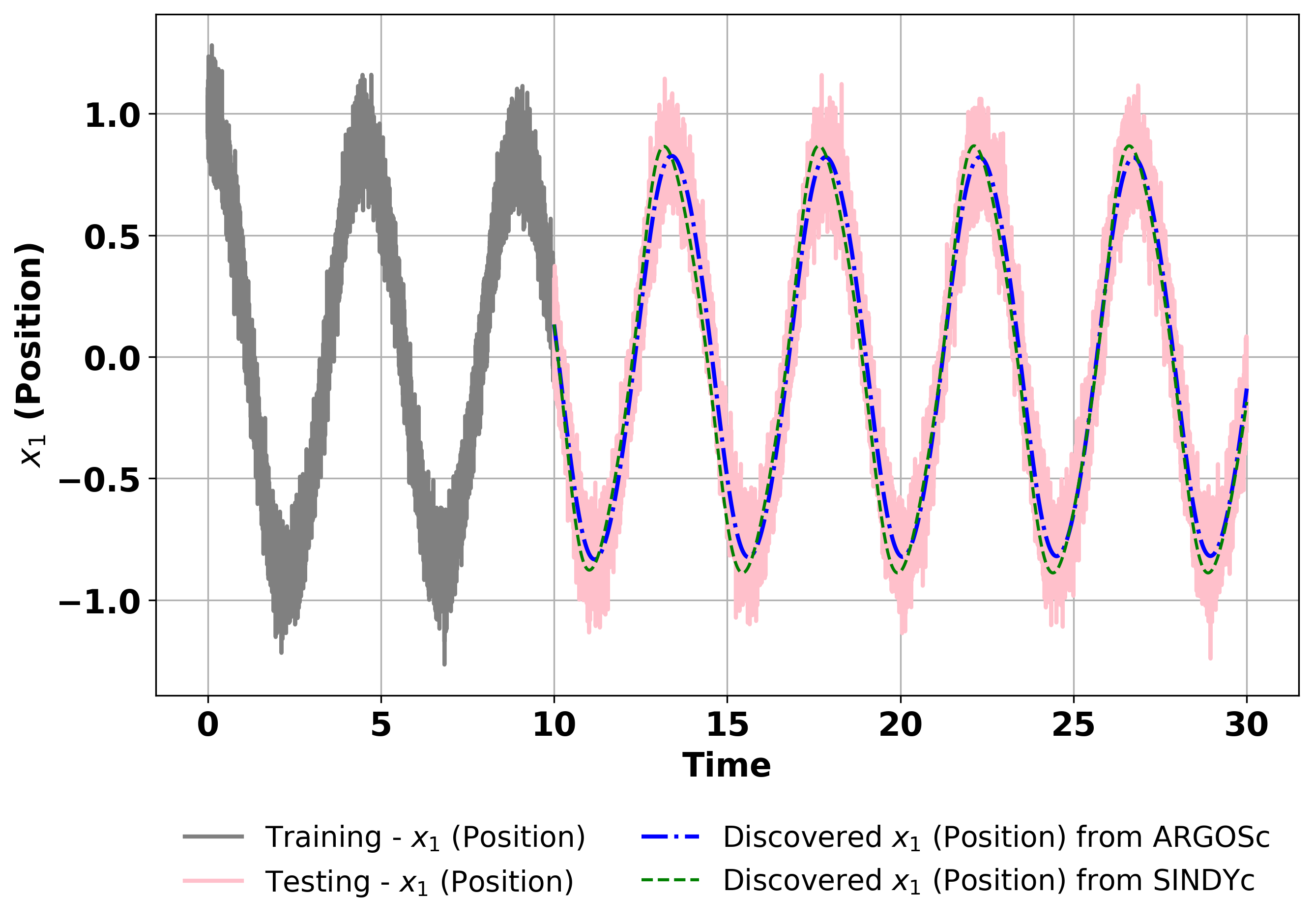}
            \label{x1_14db}
        \end{subfigure}
        \hfill
        \begin{subfigure}{0.48\textwidth}
            \centering
            \includegraphics[width=\linewidth]{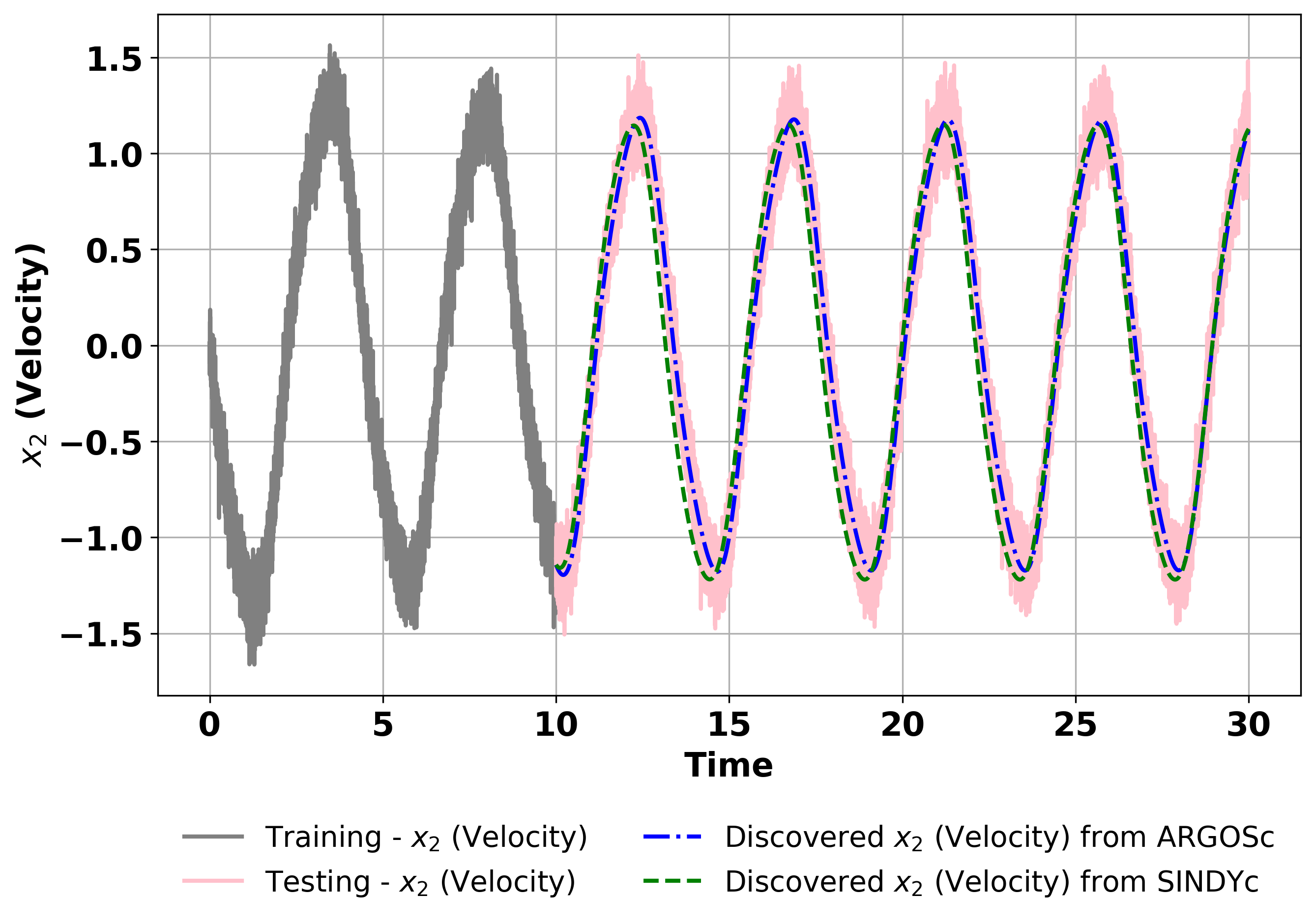}
            \label{x2_14db}
        \end{subfigure}
        \vspace{-9pt}
        \subcaption{SNR = 14 dB}
    \end{subfigure}
    
    \caption{ARGOSc and SINDYc model predictions for the Van der Pol oscillator dataset under different SNR conditions; (a) SNR = 25 dB, (b) SNR = 14 dB.}
    \label{vanderpol}
\end{figure}

The identified dynamics of the Van der Pol oscillator for both ARGOSc and SINDYc are demonstrated in Fig.\ref{vanderpol}. As shown in Fig.\ref{vanderpol}, ARGOSc successfully identified and tracked all trajectories in the testing dataset despite increasing noise due to its integration of the Savitzky-Golay filter, which effectively mitigates noise-induced distortions. In contrast, SINDYc's performance deteriorated as noise levels increased in the testing dataset. While SINDYc could still follow the general system trajectories, it was less robust than ARGOSc. Notably, both methods struggled to distinguish the intrinsic system dynamics from the feedback control laws, as the latter share similar characteristics with the underlying dynamics.

\subsection{Lotka-Volterra System} 
\begin{figure}
    \centering
    \begin{subfigure}[b]{0.5\textwidth}
        \centering
        \begin{subfigure}{0.48\textwidth}
            \centering
            \includegraphics[width=\linewidth]{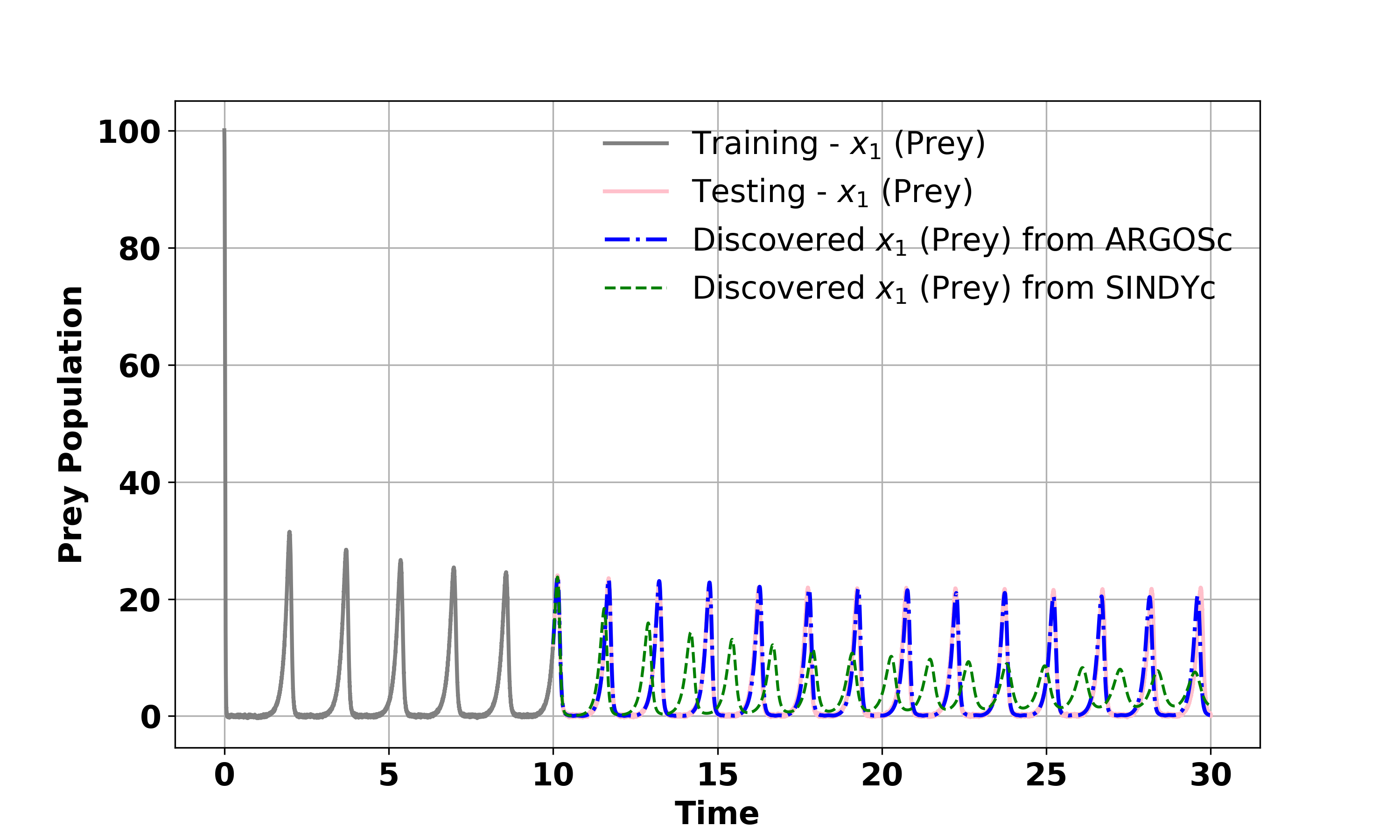}
        \end{subfigure}
        \hfill
        \begin{subfigure}{0.48\textwidth}
            \centering
            \includegraphics[width=\linewidth]{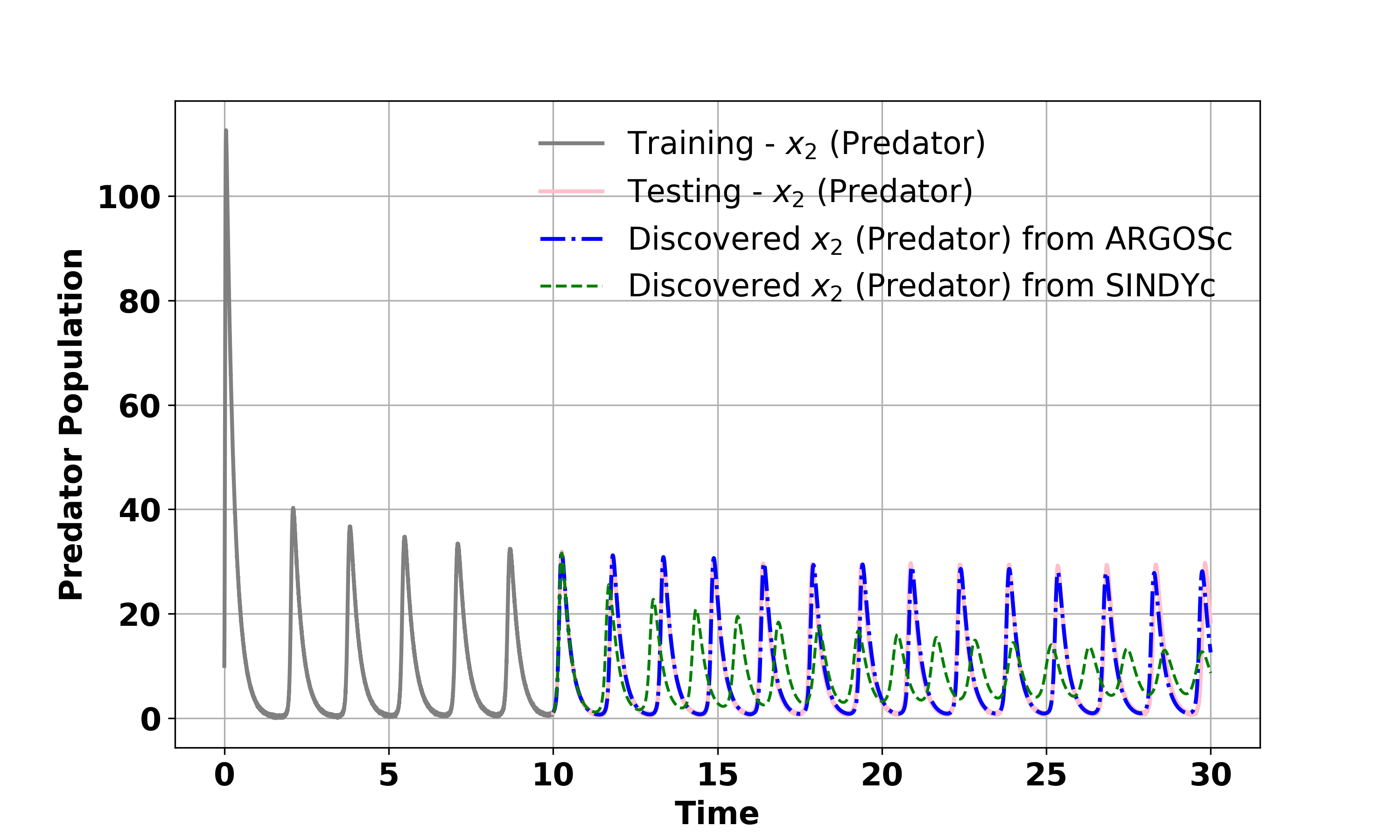}
        \end{subfigure}
        \vspace{-6pt}
        \subcaption{SNR = 25 dB}
    \end{subfigure}
    
    \begin{subfigure}[b]{0.5\textwidth}
        \centering
        \begin{subfigure}{0.48\textwidth}
            \centering
            \includegraphics[width=\linewidth]{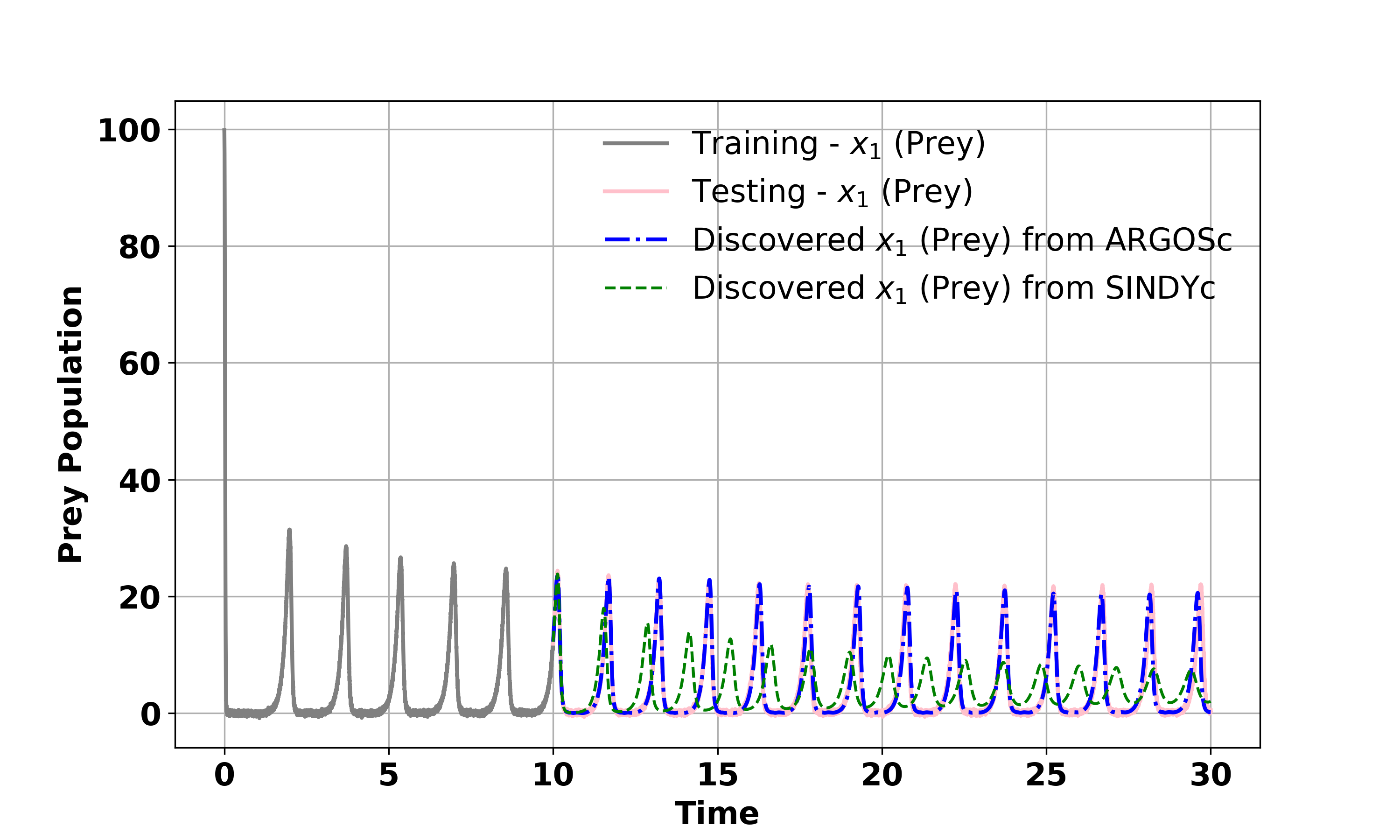}
        \end{subfigure}
        \hfill
        \begin{subfigure}{0.48\textwidth}
            \centering
            \includegraphics[width=\linewidth]{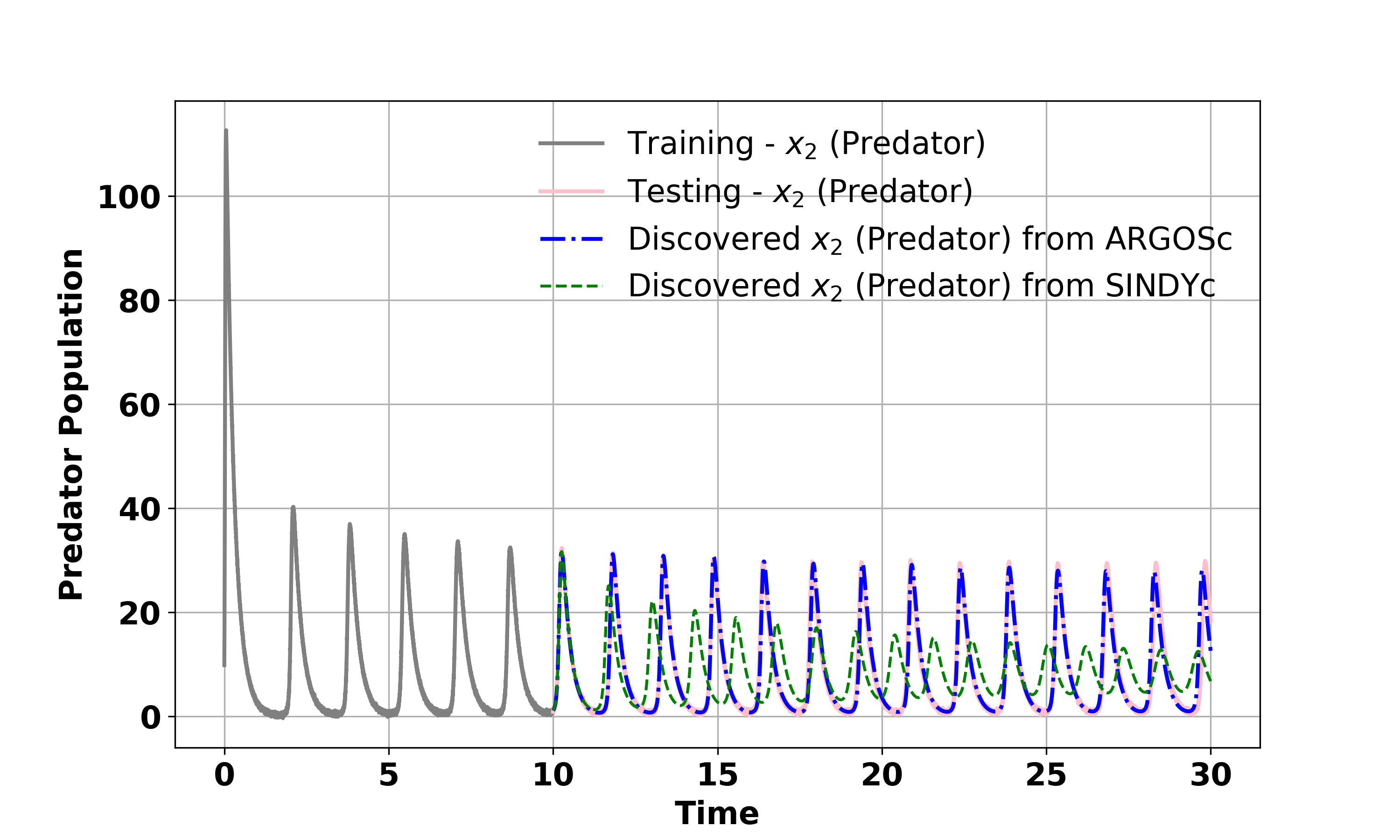}
        \end{subfigure}
        \vspace{-6pt}
        \subcaption{SNR = 14 dB}
    \end{subfigure}
    
    \caption{ARGOSc and SINDYc model predictions for  the Lotka-Volterra dataset under different SNR conditions; (a) SNR = 25 dB, (b) SNR = 14 dB.}
    \label{lotka}
\end{figure}

For this system, incorporating a sinusoidal external force into the dynamical system within a medium- to low-noise environment leads to a deterioration in SINDYc's performance, as illustrated in Fig.\ref{lotka}. 
As time progresses, SINDYc's performance deteriorates. Although this method accurately captures nearly the first two cycles of the testing phase, its amplitude and accuracy degrade over time. On the other hand, ARGOSc successfully captures all trajectories of the dynamical system in the testing stage. To provide a better understanding of the performance of each method for the corresponding dynamical system, metrics namely the mean squared error (MSE) and $R^2$ score are presented in Table~\ref{table1} and Table~\ref{table2}. 

\begin{table}[h!]
\centering
\small
\scriptsize
\setlength{\tabcolsep}{1.5pt}
\renewcommand{\arraystretch}{1.5}
\caption{Performance metrics (MSE, $R^2$) for Van der Pol oscillator and Lotka-Volterra system under 14 dB and 25 dB noise levels for $x_1$ and $x_2$.}
\resizebox{\columnwidth}{!}{
\begin{tabular}{|c|c|c|c|c|c|c|c|c|c|}
\hline
\multirow{3}{*}{\shortstack{\textbf{Dynamical}\\\textbf{System}}} & 
\multirow{3}{*}{\textbf{Method}} & 
\multicolumn{4}{c|}{\textbf{14 dB}} & 
\multicolumn{4}{c|}{\textbf{25 dB}} \\
\cline{3-10}
& & \multicolumn{2}{c|}{$x_1$} & \multicolumn{2}{c|}{$x_2$} & \multicolumn{2}{c|}{$x_1$} & \multicolumn{2}{c|}{$x_2$} \\
\cline{3-10}
& & MSE & $R^2$ & MSE & $R^2$ & MSE & $R^2$ & MSE & $R^2$ \\
\hline
\multirow{2}{*}{\shortstack{\text{Van der Pol}\\\text{oscillator}}} & SINDYc &   0.134   &   0.947   &  0.205    &   0.938   &   0.056   &   0.983   &   0.067   &  0.974    \\ \cline{2-10}
                                        & ARGOSc &   0.018   & 0.998     &  0.042    &  0.989    &  0.003    &  0.99    &  0.004    &  0.993    \\ \hline
\multirow{2}{*}{\shortstack{\text{Lotka-Volterra}\\\text{system}}}         & SINDYc &   6.932   &   -0.178   &   9.679   &  -0.166    &   6.961   &   -0.188   &   9.727   &    -0.178  \\ \cline{2-10}
                                        & ARGOSc &  0.0198    &  0.985    &  0.0438    &  0.996    & 0.0199     &  0.996    &  0.0356    &  0.984    \\
\hline
\end{tabular}
}
\label{table1}
\end{table}

\subsection{Lorenz System} 
\begin{figure}[htbp]
    \centering
    \begin{subfigure}{0.49\textwidth}
        \centering
        \includegraphics[width=\linewidth]{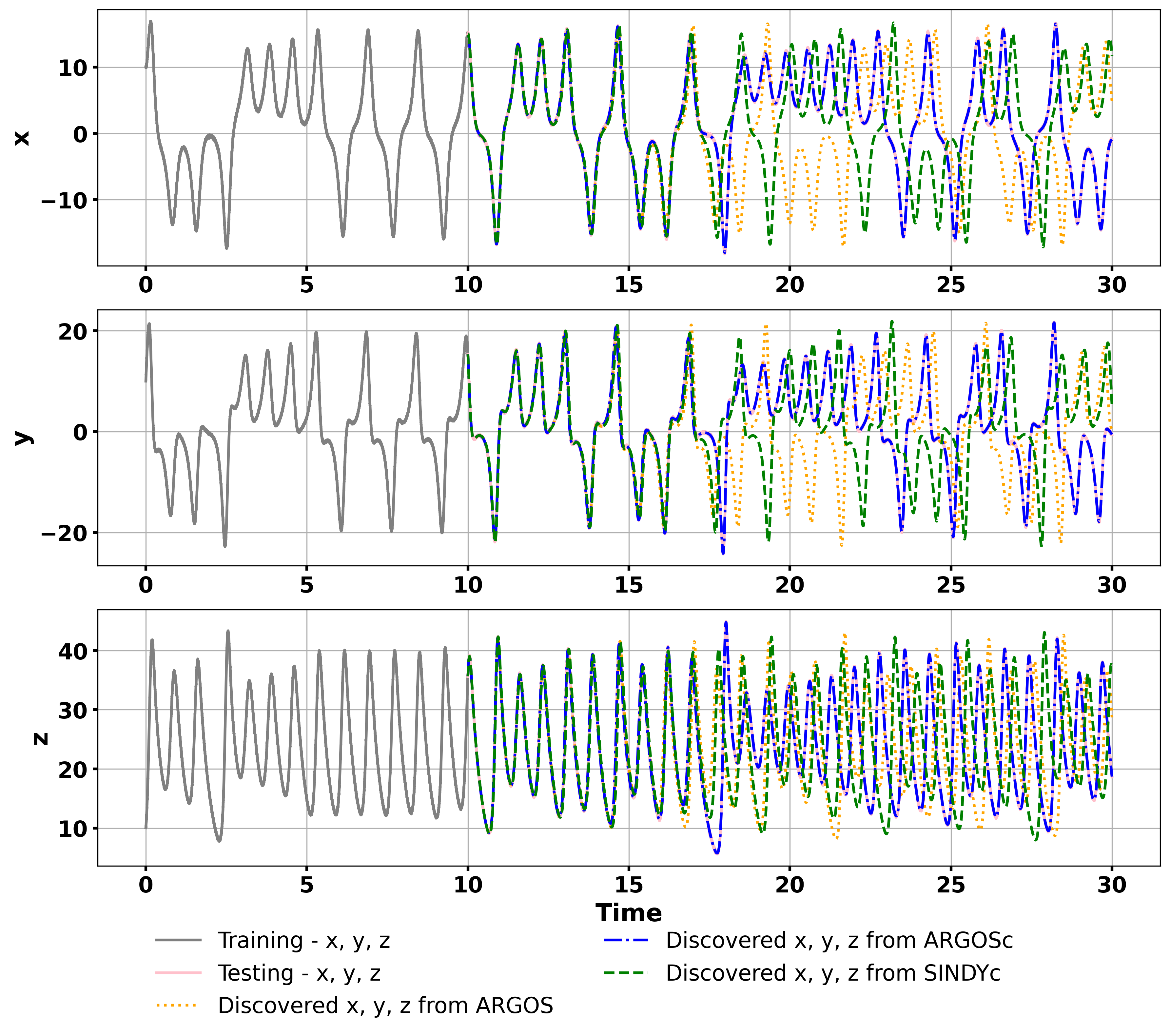}
        \caption{SNR = 49 dB}
    \end{subfigure}
    \hfill
    \begin{subfigure}{0.49\textwidth}
        \centering
        \includegraphics[width=\linewidth]{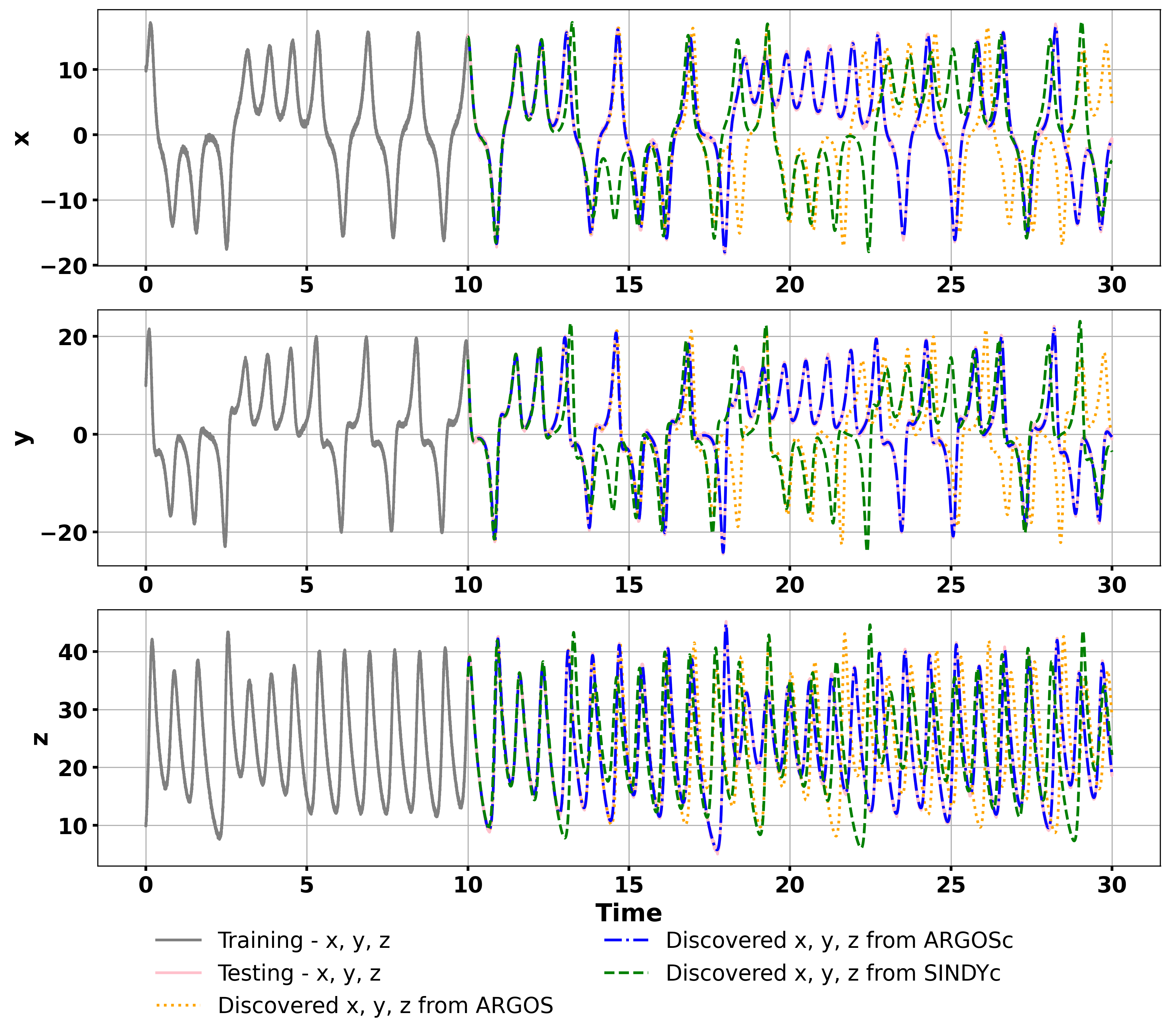}
        \caption{SNR = 37 dB}
    \end{subfigure}
    
    \caption{ARGOS, ARGOSc, and SINDYc model prediction for the Lorenz dataset under different SNR conditions; (a) SNR = 49 dB, (b) SNR = 37 dB.}
    \label{lorenz}
\end{figure}

The most challenging case study, with higher dimensionality than the previous examples, is the Lorenz system—a classic example of a chaotic system. Controlling such systems is particularly challenging due to their sensitivity to initial conditions and complex trajectories. Higher-order sinusoidal input ($\cos^3(t)$) is used as forcing law to drive this chaotic system. The results in Fig.\ref{lorenz} demonstrate the performance of ARGOSc, SINDYc, and ARGOS. It can be observed that ARGOSc more accurately identifies the system's behavior than SINDYc under noisy conditions. As shown in Fig.\ref{lorenz}, SINDYc can capture the first 7 to 8 seconds of the testing dataset at a noise level of 49 dB. However, as noise increases in the testing dataset (for an SNR of  37 dB), SINDYc's performance deteriorates, where it can capture only the first 3 seconds of the testing dataset. This method was also applied to the dataset to demonstrate the limitations of ARGOS alone. The naive application of ARGOS, without knowledge of the forcing input, leads to an unstable model that cannot capture all the trajectories of the testing dataset, unlike ARGOSc.

\begin{table}[h!]
\centering
\small
\setlength{\tabcolsep}{1.5pt}
\renewcommand{\arraystretch}{1.7}
\caption{Performance metrics (MSE, $R^2$) for the Lorenz system under 37 dB and 49 dB noise levels for $x$, $y$, and $z$.}
\resizebox{\columnwidth}{!}{
\begin{tabular}{|c|c|c|c|c|c|c|c|c|c|c|c|c|c|}
\hline
\multirow{3}{*}{\shortstack{\textbf{Dynamical}\\\textbf{System}}} & 
\multirow{3}{*}{\textbf{Method}} & 
\multicolumn{6}{c|}{\textbf{37 dB}} & 
\multicolumn{6}{c|}{\textbf{49 dB}} \\
\cline{3-14}
& & \multicolumn{2}{c|}{$x$} & \multicolumn{2}{c|}{$y$} & \multicolumn{2}{c|}{$z$} 
  & \multicolumn{2}{c|}{$x$} & \multicolumn{2}{c|}{$y$} & \multicolumn{2}{c|}{$z$} \\
\cline{3-14}
& & MSE & $R^2$ & MSE & $R^2$ & MSE & $R^2$ & MSE & $R^2$ & MSE & $R^2$ & MSE & $R^2$ \\
\hline
\multirow{2}{*}{\shortstack{\text{Lorenz}\\\text{System}}} 
& SINDYc  & 85.39 & -0.422 & 111.80 & -0.428 & 93.74 & -0.23 & 91.46 & -0.523 & 120.26 & -0.536 & 111.49 & -0.463  \\ \cline{2-14}
& ARGOSc  & 0.039 & 0.999 & 0.04 & 0.999 & 0.039 & 0.999 & 0.010 & 0.999 & 0.009 & 0.999 & 0.009 & 0.999  \\
\hline
\end{tabular}
}
\label{table2}
\end{table}

\section{Discussion}
One of the key strengths of the proposed ARGOSc learner is its ability to directly incorporate external inputs into the design matrix by including functions of both the state $x$ and the control input $u$, as well as cross terms between the state and the forcing input. This enables a more accurate and interpretable representation of forced dynamical systems. By augmenting the regression process with input-dependent terms, ARGOSc extends the applicability of the ARGOS framework to a broader range of real-world systems where forcing inputs play a fundamental role. Another key strength of the proposed method is its robustness to low- to medium-level noise, and its ability to handle external forces or control inputs, which are often present in real-world dynamical systems. This capability significantly enhances the applicability of ARGOSc for real-world scenarios, where external inputs and noise are common, making it well-suited for practical applications.

While ARGOSc has demonstrated promising results, there are several areas where further research could enhance its capabilities. First, the current implementation of ARGOSc relies on a predefined library of candidate basis functions, which may not capture all possible dynamics in more complex systems. Future work could explore adaptive feature selection techniques, such as recursive feature elimination or Bayesian optimization, to improve the accuracy and interpretability of the identified models. Second, the scalability of ARGOSc to higher-dimensional systems with multiple forcing inputs remains an open question. Extending ARGOSc to handle larger, more complex systems would broaden its applicability to real-world problems such as power systems, robotics, and climate modeling. Third, ARGOSc still struggles to distinguish feedback control laws that include state variables for stabilizing the dynamical system. Therefore, developing a methodology to separate feedback control from state variables could be a promising direction for future research. Fourth, one potential drawback of ARGOSc, compared to SINDYc, is its computational burden in discovering the dynamical system in the training set, which may limit its applicability in some real-world situations. Finally, integrating ARGOSc with adaptive control strategies, where control laws are adjusted in real time based on the identified system dynamics, could further enhance its utility in applications where system parameters are time-varying or uncertain.

\section{Conclusion}
This work extends the ARGOS framework to create ARGOSc, enabling it to handle dynamical systems with external forcing or control inputs and improving robustness to noise. The method expands the candidate nonlinear term library to include functions of both the system states and forcing inputs, along with their cross terms. It has been validated on several dynamical systems under various forcing and feedback controls. Since ARGOSc retains the computational foundation of ARGOS, it is expected to scale to similar problem classes. \uline{ARGOSc outperforms ARGOS, and SINDYc in identifying the underlying dynamics. Moreover, while SINDYc fails to capture the governing equations under noisy conditions, the proposed ARGOSc produces equations that closely match the true governing equations with high accuracy.}Future research will focus on refining feature selection and incorporating adaptive control strategies.

\addtolength{\textheight}{-12cm}   







\end{document}